\documentclass[fleqn,10pt]{wlscirep}

\usepackage{graphicx}
\usepackage{citesort}

\begin{document}

\title{Towards atomically precise manipulation of 2D nanostructures in the electron microscope}

\author{Toma Susi, Demie Kepaptsoglou, Yung-Chang Lin, Quentin M. Ramasse, Jannik C. Meyer, Kazu Suenaga, Jani Kotakoski\\
\vspace{9pt}
$^1$ University of Vienna, Faculty of Physics, Vienna 1090, Austria\\
$^2$ SuperSTEM Laboratory, SciTech Daresbury Campus, Daresbury WA4 4AD, United Kingdom\\
$^3$ National Institute of Advanced Industrial Science and Technology (AIST), Tsukuba 305-8565, Japan\\
\texttt{toma.susi@univie.ac.at; jani.kotakoski@univie.ac.at}}


\begin{abstract}
Despite decades of research, the ultimate goal of nanotechnology---top-down manipulation of individual atoms---has been directly achieved with only one technique: scanning probe microscopy. In this Review, we demonstrate that scanning transmission electron microscopy (STEM) is emerging as an alternative method for the direct assembly of nanostructures, with possible applications in plasmonics, quantum technologies, and materials science. Atomically precise manipulation with STEM relies on recent advances in instrumentation that have enabled non-destructive atomic-resolution imaging at lower electron energies. While momentum transfer from highly energetic electrons often leads to atom ejection, interesting dynamics can be induced when the transferable kinetic energies are comparable to bond strengths in the material. Operating in this regime, very recent experiments have revealed the potential for single-atom manipulation using the Ångström-sized electron beam. To truly enable control, however, it is vital to understand the relevant atomic-scale phenomena through accurate dynamical simulations. Although excellent agreement between experiment and theory for the specific case of atomic displacements from graphene has been recently achieved using density functional theory molecular dynamics, in many other cases quantitative accuracy remains a challenge. We provide a comprehensive reanalysis of available experimental data on beam-driven dynamics in light of the state-of-the-art in simulations, and identify important targets for improvement. Overall, the modern electron microscope has great potential to become an atom-scale fabrication platform, especially for covalently bonded 2D nanostructures. We review the developments that have made this possible, argue that graphene is an ideal starting material, and assess the main challenges moving forward.
\end{abstract}

\maketitle

\section*{Introduction: a new tool for nanotechnology}

Although based on physical principles it should be possible to assemble matter atom-by-atom, only few attempts at such manipulation have been successful. The first and so far only methods to truly achieve this goal are scanning probe microscopies (SPM), scanning tunneling microscopy (STM) in particular. STM is able to resolve the atomic structure at conducting surfaces by measuring the tunneling current to an atomically sharp tip. Voltage pulses from the tip also allow atoms to be picked up from one potential energy minimum and placed at the location of another one on a flat surface~\cite{Eigler90N}, or moved to a vacancy in a surface layer~\cite{Ebert93PRL}. Similarly, the interaction between the tip of an atomic force microscope (AFM) and surface atoms can lower migration barriers, allowing manipulation even at insulating surfaces~\cite{Custance09NN}. However, despite many impressive demonstrations such as quantum corrals~\cite{Crommie93S}, atomic-scale memories~\cite{Kalff16NN}, and designer atomic lattices~\cite{Drost17NP}, these techniques are by nature limited to relatively weakly bound surface adatoms or vacancies.

In transmission electron microscopy (TEM), on the other hand, the scattering of highly energetic electrons passing through the sample is used for imaging. The mass of electrons makes TEM fundamentally different from optical microscopies: since electrons carry significant momentum, they can eject atoms from the structure being observed. Although the quantum mechanical scattering of relativistic electrons from the Coulomb potential of a nucleus is a complicated process in its fine details, the ejection event can be adequately described as an elastic collision between one imaging electron and one nucleus of the target material. This electron beam damage is typically a detrimental side-effect, prompting a push towards lower acceleration voltages to avoid degrading or destroying conducting samples for which this is the dominant damage mechanism. However, atom-number-conserving changes can be stochastically induced when the transferable kinetic energies are comparable to bond strengths in the material. A crucial advantage compared to SPM is that these energies are on the order of ten electronvolts in light-atom materials even for modest electron acceleration voltages. This potentially allows atoms with strong covalent bonds to be manipulated.

Since the entire field of view is uniformly irradiated in TEM, it is difficult to control the dose at the atomic level. However, in scanning transmission electron microscopy (STEM), pre-specimen condenser lenses are used to focus the electrons into an atomically small beam, which is rastered over the sample while the scattered intensity is recorded to form an image. In modern aberration-corrected instruments, the size of the beam spot (full-width at half-maximum) is only on the order of 1 \AA~\cite{Krivanek99U,Krivanek08UM}, making it possible to essentially direct the dose at individual atomic columns. These developments have given STEM the technological maturity to become a promising candidate for a fundamentally different tool for atomically precise manipulation. Indeed, some of us recently published the very first experimental demonstration of this process~\cite{Susi17UM}.

\section*{Scanning transmission electron microscopy: enabling developments}

While STM has allowed routine atomic resolution imaging since the 1980s, the non-destructive imaging of individual atoms by STEM only became widespread recently. The challenge  was a pernicious property of electron optical systems, whereby no rotationally-symmetric electron lens of the type commonly used to form atomic-sized probes can eliminate spherical and chromatic aberrations~\cite{Hawkes09PTRSL}. The development of aberration correctors in the 1990s finally overcame this technical barrier, and electron microscopy instrumentation has seen enormous advances over the last two decades~\cite{Krivanek97Cs,Haider98N,Krivanek99U,Batson2002,Sawada2005,Krivanek08UM,Dahmen2009,Kaiser11U,Hosokawa2013}, making atomic resolution possible in most materials~\cite{Tanaka2008,Pennycook2011a}. Further, the identification of the exact chemical structure is also possible, especially in low-dimensional materials. The elemental composition can be revealed by direct Z-contrast imaging~\cite{Krivanek10N} (since the scattered intensity is proportional to the atomic number) or through energy dispersive X-ray spectroscopy~\cite{lovejoy_single_2012}, whereas electron energy loss spectroscopy (EELS)~\cite{Varela2004, Muller2008, Suenaga09NC, Colliex2012a, Bangert13NL} can additionally reveal the exact bonding configuration. Non-destructive imaging of a range of materials has meanwhile been enabled by lower electron acceleration voltages~\cite{Suenaga09NC,Krivanek10N,Kaiser11U}, yet still retaining atomic resolution.

Early aberration-corrected instruments were, however, too sensitive to focus the electron beam on a single atom for extended periods of time, and typical residual vacuum pressures in the instrument columns led to chemical etching~\cite{Meyer12PRL}. Sample stability is very important for EELS mapping, but particularly crucial for single-atom manipulation where the electron dose needs to be accurately directed. These challenges have been largely addressed in the latest generation of instruments, leading to remarkable sample stability. Stage drift can be as low as 0.1~{\AA} per minute, and, thanks to improved vacuum, beam damage in pristine graphene is completely suppressed at electron acceleration voltages below 80 kV~\cite{Meyer12PRL,Susi16NC}.

\section*{Graphene: advantage of two dimensions}

Concurrently with improvements in STEM, the discovery of graphene~\cite{Novoselov05N} has provided ideal samples for atomic resolution studies~\cite{Meyer08NL}. Not only is monolayer graphene highly conductive, which effectively mitigates other damage mechanisms such as radiolysis and ionization, it also has remarkable electronic and mechanical properties~\cite{Geim07NM}. These are theoretically well understood, and possible to accurately model from first principles using modern computational techniques.

The greatest advantage of graphene for manipulation experiments is its reduced dimensionality. The two-dimensional atomic structure can be directly imaged and atomic species easily identified, both of the material itself as well as any present impurities~\cite{Krivanek10N}. Even in cases where a projected image may be ambiguous, as when a heavier impurity atom such as Si~\cite{Ramasse13NL,Zhou12PRL,Susi14PRL} or P~\cite{Susi172DM} buckles out of the plane, EELS can be used to unambiguously identify the precise nature of the bonding. Further, the electron dose can be predominantly directed at single atoms, apart from probe tails, \textit{i.e.} the spreading of the intensity envelope of the electron beam beyond its sharp maximum~\cite{Kotakoski14NC}. Aberration-corrected beams are so finely focused that if the dose is directed at a single carbon atom in graphene, we estimate that only 0.3\% impinges upon the other two neighbors of an impurity (integrated beam intensity based on the electron probe shape measured in~\cite{Kotakoski14NC}). Since the relevant atom-scale dynamics are caused by electron impacts on a specific carbon atom~\cite{Susi14PRL}, this precision is crucial for control.

The reduced dimensionality also considerably simplifies both the physical interaction between the beam and the sample, and the analysis of the recorded image data. Each electron interacts only with a quasi-2D electron density, and predominantly scatters from the electrostatic potential of a single atomic nucleus. Besides simplifying theoretical descriptions of the interaction, the atomic structure can thus be directly and unambiguously identified from the scattered signal. This stands in stark contrast to three-dimensional crystals, where the complex interactions between the probe electrons and the sample, and the precise real-time reconstruction of its three-dimensional lattice from projected images present daunting challenges for atomically precise control~\cite{Kalinin16N}.

\section*{Atom-scale dynamics: state-of-the-art}

In addition to spatial resolution, understanding dynamics requires collecting information in the time domain. Atomic motion that happens on femto-to-picosecond timescales is several orders of magnitude too fast to capture using conventional instruments, and even dedicated ultrafast techniques are limited to nanosecond timescales for real-space imaging far from atomic resolution~\cite{Plemmons15CM, Vogelsang15NL}. However, if the probability (\textit{i.e.} cross section) of an electron-beam-driven process is low enough compared to the irradiation dose rate, each event can be individually distinguished. This enables their statistical treatment as a Poisson process~\cite{Susi14PRL}, allowing experimental interaction cross sections to be extracted and compared to simulations.

In the case of graphene, there is a growing body of work on the controlled creation~\cite{Robertson12NC} and annihilation~\cite{Kurasch12NL} of defects, and etching and edge shaping directed by the electron beam~\cite{Girit09S,Lin14NN}. However, our focus here is on non-destructive (atom-number-conserving) manipulation. Fundamentally, such reversible beam-driven processes in graphene can be divided into two categories: bond rotation events and impurity dynamics.

\begin{figure}[h]
\centering
\includegraphics[width=1.0\linewidth]{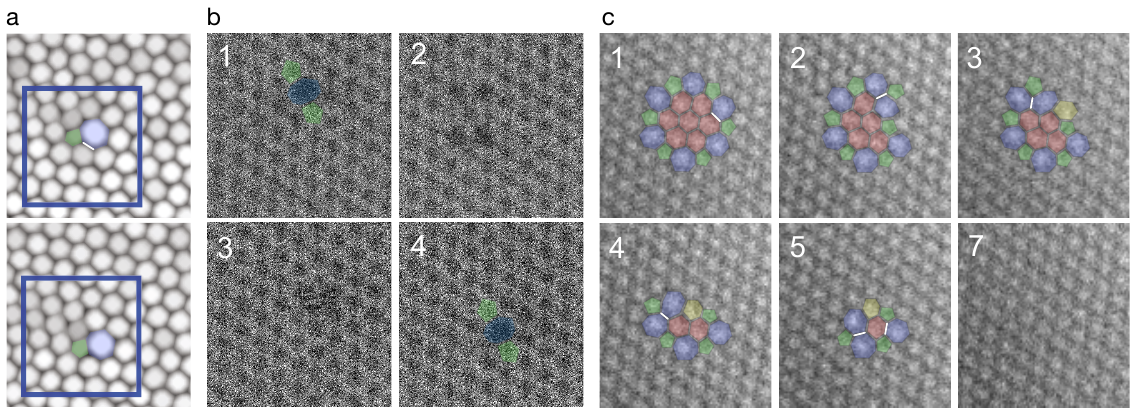}
\caption{\textbf{Bond rotation processes in graphene} (selected frames with fields of view $\sim$2.5 nm, rotating bonds highlighted in white, polygons in non-pristine graphene areas overlaid in color). a) Dislocation glide (TEM, adapted with permission from ~\cite{Lehtinen13NC}. Copyright 2013, Rights Managed by Nature Publishing Group), b) Four frames of a divacancy random walk with several (unmarked) bond rotations in between~\cite{Kotakoski14NC} (STEM), c) Flower defect annihilation through one bond rotation at a time (TEM, reprinted with permission from~\cite{Kurasch12NL}. Copyright 2012 American Chemical Society).}
\label{fig:rotation}
\end{figure}

Most of the former observations have been of all-carbon structures, ranging from the creation and annihilation of the double pentagon-heptagon defect (controversially~\cite{Monthioux14Carbon} called "Stone-Wales defect" in the current literature) in pristine graphene~\cite{Wakabayashi07NN,Meyer08NL} and bond rotations at a graphene edge~\cite{Gong14ACSNano} to the migration of dislocation cores~\cite{Warner12Science,Lehtinen13NC} and divacancies~\cite{Kotakoski11PRL,Kotakoski14NC,Borner16PRB}, and the transformation of grain boundaries~\cite{Kurasch12NL}. The bond rotation process is also responsible for the healing of the so-called flower defect under electron irradiation~\cite{Kurasch12NL}. Crucially, in all of these cases, the structural changes in the material can be explained through rotated carbon-carbon bonds, understood to occur due to a single impact from an electron to one of the involved carbon atoms~\cite{Kotakoski11PRB}. In Fig.~\ref{fig:rotation}, we present examples for some of these cases, pointing out the individual bonds that have rotated between the frames when possible.

\begin{figure}[h]
\centering
\includegraphics[width=1.0\linewidth]{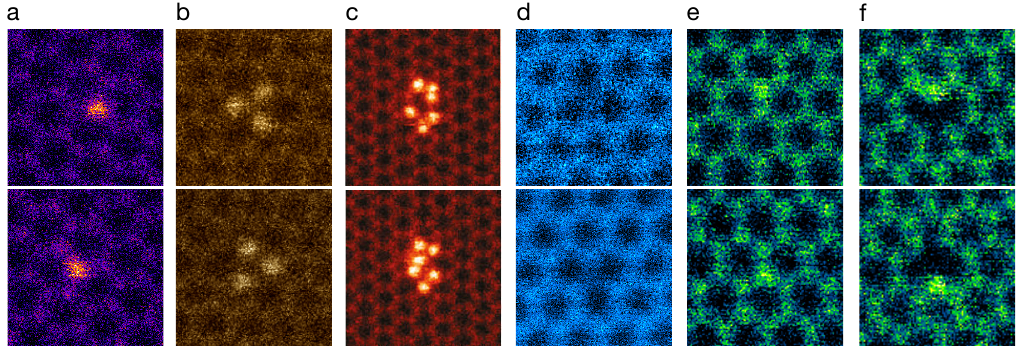}
\caption{\textbf{Atom-number-conserving impurity dynamics in graphene} (STEM, field of view $\sim 1 \times 1$ nm; panel c $\sim 2\times 2$ nm). a) Si--C bond inversion at a trivalent Si substitution (data from published experiments~\cite{Susi14PRL}), b) Rotation of Si trimer in a divacancy (data from supplementary information of~\cite{Yang14AC}), c) Si atom movement within a Si$_6$ cluster in a pore (reproduced with permission~\cite{Lee13NC}, Copyright \copyright 2013, Rights Managed by Nature Publishing Group), d) B--C bond inversion at a trivalent B substitution (data from published experiments~\cite{Ramasse13NL,Kepaptsoglou15AN}), e) N--C bond inversion at a trivalent N impurity (data from published experiments~\cite{Lin15NL}). f) Bivalent N jumps across a vacancy (data from published experiments~\cite{Lin15NL}).}
\label{fig:impurity}
\end{figure}

In contrast to observations of bond rotations, only a handful of examples of beam-driven and non-destructive impurity motion in graphene are known. These are the movement of the trivalent Si substitution via an out-of-plane bond inversion process~\cite{Susi14PRL}, the rotation of a Si trimer in a divacancy~\cite{Yang14AC}, the atomic motions in a Si$_6$ cluster in a pore~\cite{Lee13NC}, and the jumping of a bivalent N between two equivalent bonding sites across a single vacancy~\cite{Lin15NL}. Although not previously discussed, we have observed both B and N substitutions to also undergo lattice jumps similar to Si (data from published experiments~\cite{Ramasse13NL,Kepaptsoglou15AN,Lin15NL}). Fig.~\ref{fig:impurity} illustrates each of these processes, with the impurity atoms appearing brighter than the carbon atoms of the lattice due to their higher atomic number (apart from B, which appears dimmer). In the case of the trivalent Si impurity, a physical mechanism to direct the atom-scale dynamics is known~\cite{Susi14PRL}, pointing a way for atom manipulation experiments.

\section*{Electron irradiation: physics of manipulation}

The highly energetic electron can transfer a maximal amount of kinetic energy when it backscatters from the electrostatic potential of an atomic nucleus. For typical relativistic velocities this process occurs on the 10$^{-21}$ s time scale, allowing it to be treated as an instantaneous event even compared to electron dynamics. When the transferred energy is larger than what is required to remove an atom from the immediate vicinity of its lattice site (called the displacement threshold), knock-on damage occurs.

At room temperature, pristine graphene is damaged~\cite{Meyer12PRL,Susi16NC} at electron acceleration voltages above 80 kV (corresponding to an experimentally estimated~\cite{Susi16NC} displacement threshold of 21.14 eV), even leading to the amorphization of the material~\cite{Kotakoski11PRL} during extended experiments at 100 kV. Excitations in highly conductive graphene are damped extremely fast, and each impact can thus be accurately described as a perturbation of the equilibrium state~\cite{Susi16NC}. How well this assumption holds for imperfect structures needs to be confirmed, but early indications show a puzzling discrepancy with theoretical expectations for the rate of bond rotations under 60 kV TEM observation~\cite{Skowron16C}, as well as at impurity sites as we discuss below.

\begin{figure}[h]
\centering
\includegraphics[width=0.8\linewidth]{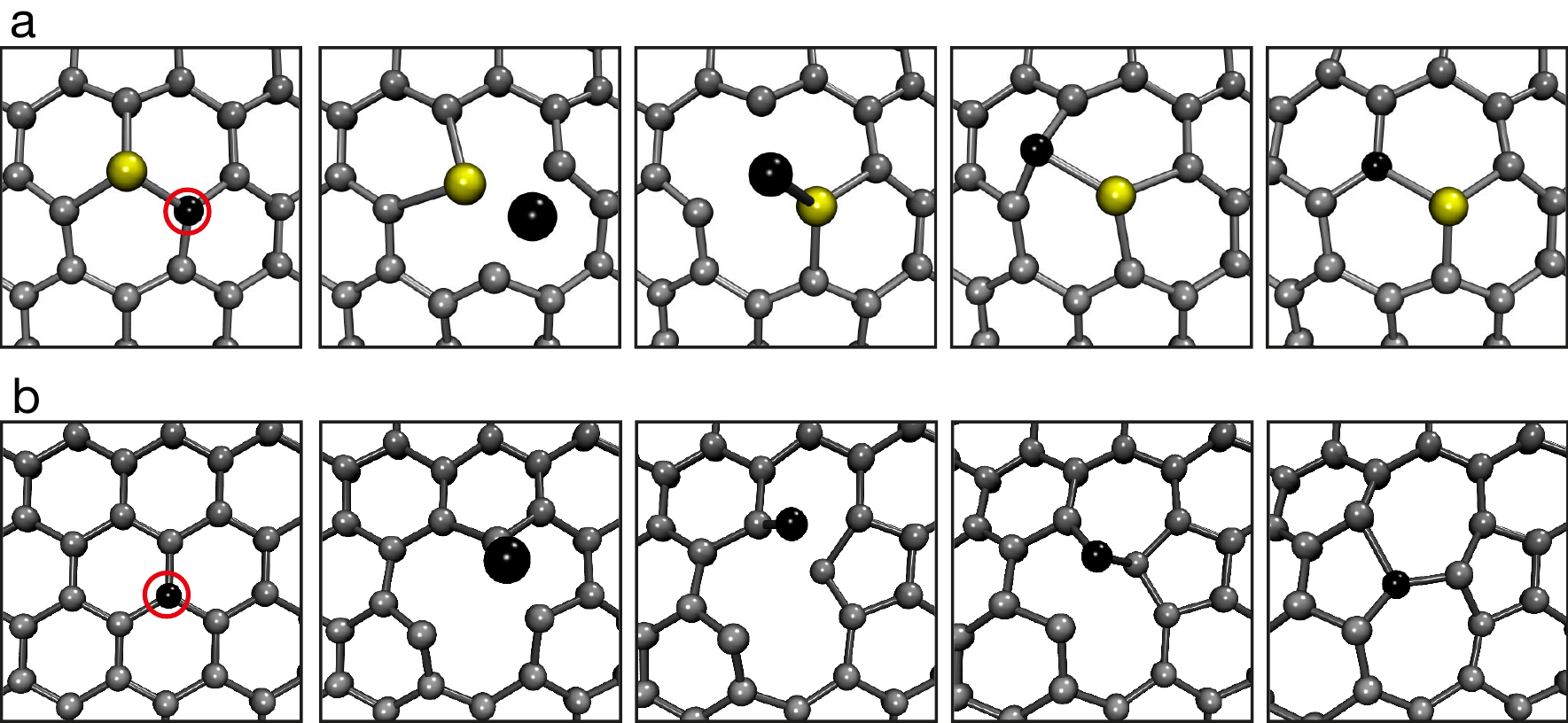}
\caption{\textbf{Mechanisms for single-atom manipulation in graphene revealed by modeling.} The carbon atom undergoing out-of-plane dynamics caused by the nuclear backscattering of a single electron is colored black (and the silicon in panel a, yellow). By focusing the electron irradiation on the desired atom (denoted by the red circle in the first frame), it is possible to direct the dynamics. a) The silicon--carbon bond inversion (simulation published in~\cite{Susi14PRL}). b) The Stone-Wales bond rotation (momentum transfer at an 12.5$^{\circ}$ angle, simulation published in~\cite{Kotakoski11PRB}).}
\label{fig:mechanism}
\end{figure}

For atomic-scale manipulation of structures with impurities, mass is important since heavier heteroatoms receive less energy in a momentum-conserving interaction with an electron. However, carbon atoms next to an impurity tend to be more weakly bound than atoms of the bulk, leading to their preferential ejection~\cite{Susi12AN}. This effect can impose a direction for atomic motion caused by sub-displacement electron impacts. For three-coordinated Si impurities in graphene irradiated by 60-keV electrons, these considerations result in a striking effect~\cite{Susi14PRL}. Although electrons of such energy are unlikely to cause outright damage, they can momentarily displace the C neighbor of Si. In the resulting dynamical out-of-plane process revealed by modeling~\cite{Susi14PRL} (Fig.~\ref{fig:mechanism}a), the Si relaxes into the vacancy created by the impact on C, which is pulled back into the lattice on the opposite side. This makes the switching of places of the Si and C---reminiscent of flipping a digital bit from one state to the other one---highly directional and enables control over it~\cite{Susi14PRL,Susi17UM}. In the case of bond rotations, the process is very similar~\cite{Kotakoski11PRB} (Fig.~\ref{fig:mechanism}b). For bonds in defective graphene structures~\cite{Kotakoski14NC}, rotations that lead to rare atomic configurations (such as carbon rings with four or fewer atoms) are less likely to occur than those involving only pentagons, hexagons and heptagons. Hence, it should be possible to control the atomic structure by placing the electron beam on top of the bond one wishes to rotate to maximize the electron dose on the two involved atoms. To our knowledge, control of bond rotations has not been attempted. However, we recently showed that by iteratively placing the electron probe on top of a selected carbon neighbor, it is indeed possible to "pull" a silicon impurity through the graphene lattice~\cite{Susi17UM}.

\section*{First principles modeling: pushing the limits}

Modeling atomic dynamics accurately over the picosecond timescales required to extract displacement thresholds is demanding. Most computationally affordable tight-binding (TB) models have failed to yield accurate thresholds for 2D materials with non-carbon atoms, necessitating the use of more expensive density functional theory (DFT) for dynamical simulations (e.g. N-doped graphene~\cite{Loponen06PRB,Susi12AN}, hexagonal boron nitride (hBN)~\cite{Zobelli07PRB,Kotakoski10PRB}). These studies have established DFT-based molecular dynamics (DFT/MD) as the most reliable way to theoretically estimate the displacement threshold energies.

In addition, at electron energies near the threshold, the vibration of nuclei in the direction of the beam become important in activating otherwise energetically prohibited processes~\cite{Meyer12PRL,Susi16NC}. By modeling the atomic motion via a quantum description of lattice vibrations, it is possible to estimate the cross section of the knock-on process. When the velocity distribution in the perpendicular direction was included in the model, the probability of displacements from $^{12}$C and $^{13}$C graphene could be directly predicted from theory~\cite{Susi16NC}. However, all of the tested exchange correlation functionals (including some with a description of the van der Waals interaction) seem to  overestimate the displacement threshold, with the closest match within 0.3 eV (1.4\%) of the best-fit experimental threshold. Although this agreement is good enough that the DFT-derived displacement cross section values are within experimental uncertainties, this is achieved by selecting the functional providing the closest match rather than the physically best motivated theoretical model. The explicit phonon calculations required are also only feasible for small numbers of atoms, making it difficult to extend precise analyses to systems with impurities. For their displacement, discussed below, we thus merely correct the mass term in the mean square velocity with the impurity mass.

With these caveats in mind, in Table~\ref{tab:thresholds} we show our reanalysis of published data using the latest out-of-plane vibration model~\cite{Susi16NC}. For the case of N~\cite{Susi12AN} and Si~\cite{Susi14PRL}, we find discrepancies of 1.7 and 4.1 eV in the simulated thresholds for the ejection of C atoms neighboring the impurity. Using data collected on B during the experiments described in~\cite{Kepaptsoglou15AN}, we find discrepancies of 2.1 and 4.5 eV for B and C ejection, with the theoretical thresholds leading to negligible displacement rates. Calculating cross sections with a 50\% increased mean square velocity of vibrations would reduce the discrepancies, but even such a high correction does not bring the consistently underestimated theoretical cross section values into agreement with experiment. 

Finally, we have occasionally observed the ejection of the N impurity itself upon intense spot irradiation at 60 keV, hinting at the possibility of dose-rate dependence at impurity sites that is not included in the modeling, or to the occurrence of rare chemical etching events. The most serious discrepancy with current theory, however, results from the observed rate of reversible jumps of a pyridinic N dopant across a vacancy~\cite{Lin15NL}. The thermal barrier for such a transformation is at least 4 eV~\cite{Arenal14NL}, leaving beam activation as the only plausible mechanism even in high temperature experiments (we have once also observed the same event at room temperature). The experimental rate calculated from seven consecutive jumps corresponds to a cross section of over 30 barn, whereas DFT/MD modeling fails to reproduce the event (the jump is only initiated for in-plane momentum transfers that are unphysically large for the experimental geometry), let alone quantify its probability.

The physical reasons behind such differences need to be understood and the theoretical treatments correspondingly improved to provide accurate predictions for manipulation experiments. Apart from neglecting vibrational impurity modes, the remaining inaccuracy of DFT/MD could be due to one or more of the presently necessary approximations: 1) the system size is limited due to the computational cost; 2) the simulation timestep is likewise limited, possibly leading to cumulative errors in the integrated equations of motion~\cite{Lippert07JCP}; and 3) the description of the dynamics may be inadequate either due to the approximation of exchange and correlation, or the lack of time-dependence of the electronic degrees of freedom.

Although the out-of-plane buckling of silicon~\cite{Ramasse13NL} (or phosphorus~\cite{Susi172DM}) provides a strong asymmetry for the atomic motion even for momentum transfers completely perpendicular to the plane, for impurities with planar bonding geometries such as N and B, in-plane components of the momentum either from the electron impact or from phonons (or contraction of bonds at the moment of the impact due to the vibrations) may be required to explain the observed dynamics. A coordinated experimental and theoretical effort should be able to bring light on these issues.

A related but different topic where first-principles or other type of quantitative modeling has not even been attempted is radiation damage by ionization or electronic excitations (also known as radiolysis)~\cite{Egerton10UM}. Although irradiation effects in graphene seem to be adequately described solely by the knock-on effect~\cite{Susi16NC}, as the materials zoo is extended---especially if soft matter and molecular structures are included---radiolysis becomes important. Ionization damage is already the most obvious hindrance for manipulation of non-graphene materials, and our understanding of radiation damage is inevitably incomplete without its quantitative understanding. Isotope-labeled hBN or transition metal dichalcogenides would make for excellent systems to isolate radiolysis from knock-on damage, especially when supplemented with experiments combining these materials with graphene to mitigate ionization~\cite{Algara-Siller13APL,lehnert_electron_2017}.

\section*{Outlook: paths forward}

\begin{figure}[h]
\centering
\includegraphics[width=1.0\linewidth]{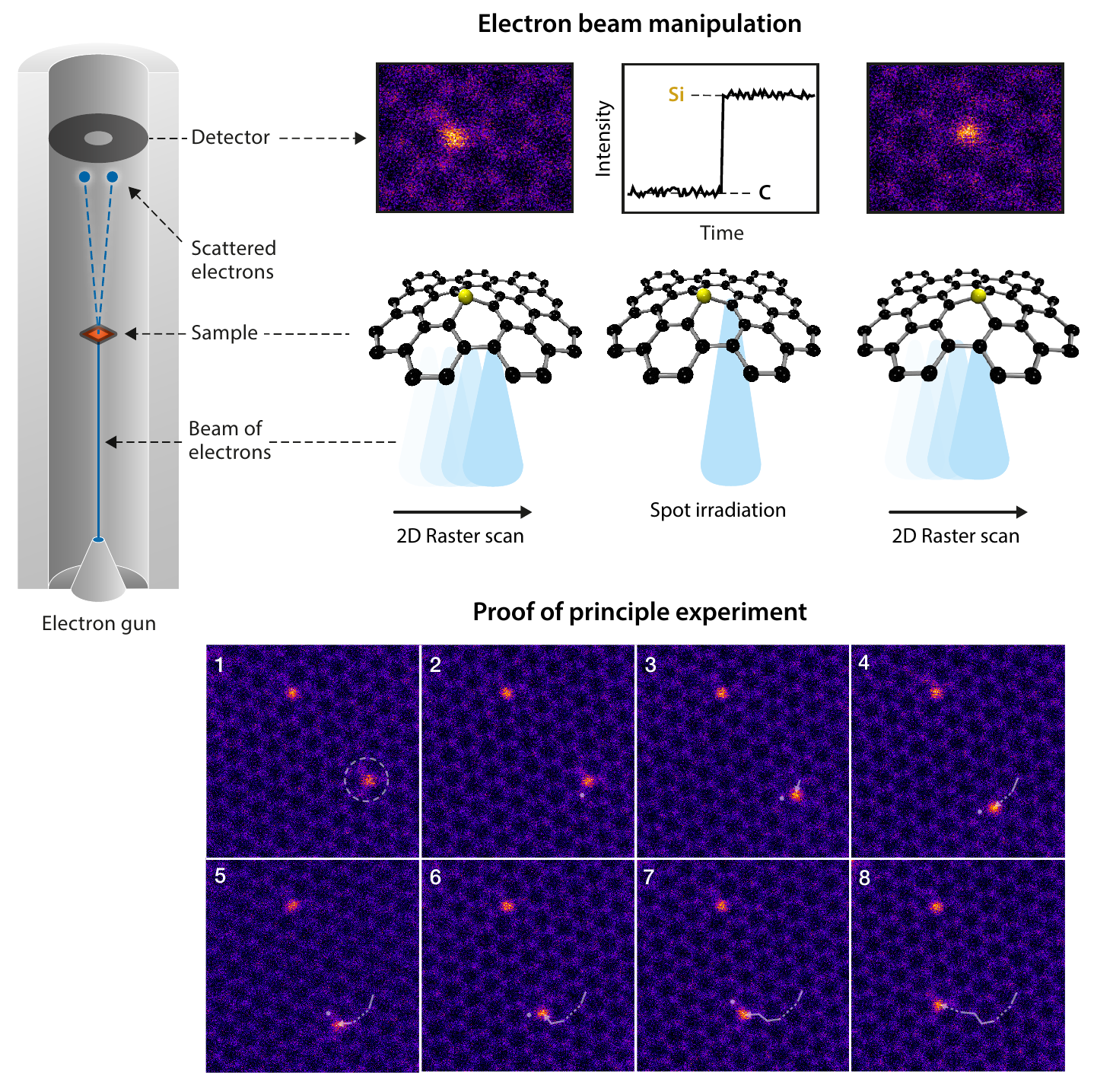}
\caption{\textbf{STEM electron beam manipulation of Si impurities in graphene.} \textbf{Top:} Schematic illustration of the manipulation process. An important step for automation would be the automatic readout of the scattered intensity during spot irradiation (here illustrating the sudden increase in intensity when the heavier Si atom moves to the irradiated lattice site). \textbf{Bottom}: Proof-of-principle experiment (MAADF detector, 60 kV, time between frames 15 s). In the first frame, the three-coordinated Si impurity selected for manipulation is denoted by the dashed circle. In the following frames, the solid dots denote the location of the manually parked electron beam, the solid arrows the cumulative single-site jumps, and dashed arrows jumps over more than one site. (Adapted with permission from~\cite{Susi17UM}, Copyright 2017 Elsevier B.V.)}
\label{fig:automove}
\end{figure}

To develop STEM into a practical manipulation technique, in addition to better modeling of beam-induced dynamics, improvements in sample preparation and the automation of manipulations are required. In our view, Si impurities in graphene provide the most promising system for initial experiments, as shown in Fig.~\ref{fig:automove}. In this case, the process of moving the Si atom in graphene was carried out by manually directing the beam. While this is how also the first successes of STM were achieved, it is impractical to fabricate complex nanostructures in such a manner. Fortunately, the modern computerized STEM is well-suited for automation, and the required software tools can be swiftly developed, along the lines discussed in~\cite{Kalinin16N,Susi17UM}.

Sample quality presents another obstacle for large-scale manipulation. Even a single atomic layer of contamination prevents such experiments, and finding clean areas, especially those that contain heteroatoms~\cite{Bangert13NL,Susi172DM}, is difficult. Even when such areas are found, contamination often gathers under the beam and obscures the surface~\cite{Meyer08APL}. Finally, a sufficient quantity of heteroatoms need to be introduced into the lattice in the first place, but without causing significant structural damage or contamination. Ion implantation is a particularly promising technique, but contamination remains a serious issue~\cite{Bangert13NL,Susi172DM}. Cleaning the samples outside the microscope vacuum~\cite{Lin12NL,Algara-Siller14APL} seems insufficient for ideal samples, since atmospheric exposure will be particularly detrimental to the chemically more reactive impurity sites. \textit{In situ} heating of samples appears more promising, either in a dedicated sample holder with a resistive heating element~\cite{Lin15NL}, via Joule heating~\cite{Lu11NL} or using a laser~\cite{Tripathi17PSSRRL}.

Current literature almost exclusively provides examples of reversible electron-beam driven dynamics in graphene, largely because of its availability and robustness as an electron microscopy specimen. However, there is no fundamental reason why other 2D materials~\cite{Hong17AM} could not host similar processes. If ionization damage can be mitigated, possibilities include 2D SiO$_2$~\cite{Bjorkman13SciRep}, silicene~\cite{Gao13Nanoscale} and phosphorene~\cite{Hu15JPCC}, which all share bond-rotation-type defects with graphene. Other candidates could be directed vacancy or impurity atom migration in transition metal dichalcogenides~\cite{Komsa12PRL}.

Finally, although two-dimensional crystals are ideal for manipulation due to their relative simplicity both experimentally and theoretically, the penetration of an electron beam through a sample offers the possibility of bulk manipulation. The methods and tools developed in 2D would largely translate to materials more generally, allowing us to eventually tackle the full complexities of three-dimensional crystals~\cite{Kalinin16N}.

\section*{Conclusion}
The challenges that we have discussed above are significant, but so is the current rate of progress. The rewards, however, are even greater: all properties of a material are determined by its chemical structure, whose precise control would allow these to be designed at will within the bounds given only by the laws of physics and material stability. Electron beam manipulation will in the near future allow the creation and tailoring of covalently bonded 2D nanostructures, with a further possibility of extending the technique to 3D. Initial experiments could for example target designed molecules embedded within graphene~\cite{Guo14NC}, plasmonic nanoantennas~\cite{Zhou12NN}, and novel quantum corrals~\cite{Susi15FWF}. Once established, atomically precise manipulation in the electron microscope will open a new playground for materials science and engineering at the ultimate limit of control.

\clearpage
\bibliography{References_BibDesk,library,bondrotations}

\begin{thebibliography}{10}
\expandafter\ifx\csname url\endcsname\relax
  \def\url#1{\texttt{#1}}\fi
\expandafter\ifx\csname urlprefix\endcsname\relax\def\urlprefix{URL }\fi
\expandafter\ifx\csname doiprefix\endcsname\relax\def\doiprefix{DOI }\fi
\providecommand{\bibinfo}[2]{#2}
\providecommand{\eprint}[2][]{\url{#2}}

\bibitem{Eigler90N}
\bibinfo{author}{Eigler, D.~M.} \& \bibinfo{author}{Schweizer, E.~K.}
\newblock \bibinfo{title}{Positioning single atoms with a scanning tunnelling
  microscope}.
\newblock \emph{\bibinfo{journal}{Nature}} \textbf{\bibinfo{volume}{344}},
  \bibinfo{pages}{524--526} (\bibinfo{year}{1990}).
\newblock \doiprefix 10.1038/344524a0.

\bibitem{Ebert93PRL}
\bibinfo{author}{Ebert, P.}, \bibinfo{author}{Lagally, M.~G.} \&
  \bibinfo{author}{Urban, K.}
\newblock \bibinfo{title}{Scanning-tunneling-microscope tip-induced migration
  of vacancies on {GaP}(110)}.
\newblock \emph{\bibinfo{journal}{Phys. Rev. Lett.}}
  \textbf{\bibinfo{volume}{70}}, \bibinfo{pages}{1437--1440}
  (\bibinfo{year}{1993}).
\newblock \doiprefix 10.1103/PhysRevLett.70.1437.

\bibitem{Custance09NN}
\bibinfo{author}{Custance, O.}, \bibinfo{author}{Perez, R.} \&
  \bibinfo{author}{Morita, S.}
\newblock \bibinfo{title}{Atomic force microscopy as a tool for atom
  manipulation}.
\newblock \emph{\bibinfo{journal}{Nature Nanotechnology}}
  \textbf{\bibinfo{volume}{4}}, \bibinfo{pages}{803--810}
  (\bibinfo{year}{2009}).
\newblock \doiprefix 10.1038/nnano.2009.347.

\bibitem{Crommie93S}
\bibinfo{author}{Crommie, M.~F.}, \bibinfo{author}{Lutz, C.~P.} \&
  \bibinfo{author}{Eigler, D.~M.}
\newblock \bibinfo{title}{Confinement of electrons to quantum corrals on a
  metal surface}.
\newblock \emph{\bibinfo{journal}{Science}} \textbf{\bibinfo{volume}{262}},
  \bibinfo{pages}{218--220} (\bibinfo{year}{1993}).
\newblock \doiprefix 10.1126/science.262.5131.218.

\bibitem{Kalff16NN}
\bibinfo{author}{Kalff, F.~E.} \emph{et~al.}
\newblock \bibinfo{title}{A kilobyte rewritable atomic memory}.
\newblock \emph{\bibinfo{journal}{Nature Nanotechnology}}
  \textbf{\bibinfo{volume}{11}}, \bibinfo{pages}{926--929}
  (\bibinfo{year}{2016}).
\newblock \doiprefix 10.1038/nnano.2016.131.

\bibitem{Drost17NP}
\bibinfo{author}{Drost, R.}, \bibinfo{author}{Ojanen, T.},
  \bibinfo{author}{Harju, A.} \& \bibinfo{author}{Liljeroth, P.}
\newblock \bibinfo{title}{Topological states in engineered atomic lattices}.
\newblock \emph{\bibinfo{journal}{Nature Physics}}
  \textbf{\bibinfo{volume}{13}}, \bibinfo{pages}{668--671}
  (\bibinfo{year}{2017}).
\newblock \doiprefix 10.1038/nphys4080.

\bibitem{Krivanek99U}
\bibinfo{author}{Krivanek, O.~L.}, \bibinfo{author}{Dellby, N.} \&
  \bibinfo{author}{Lupini, A.~R.}
\newblock \bibinfo{title}{Towards sub-{{\AA}} electron beams}.
\newblock \emph{\bibinfo{journal}{Ultramicroscopy}}
  \textbf{\bibinfo{volume}{78}}, \bibinfo{pages}{1 -- 11}
  (\bibinfo{year}{1999}).
\newblock \doiprefix 10.1016/S0304-3991(99)00013-3.

\bibitem{Krivanek08UM}
\bibinfo{author}{Krivanek, O.~L.} \emph{et~al.}
\newblock \bibinfo{title}{An electron microscope for the aberration-corrected
  era}.
\newblock \emph{\bibinfo{journal}{Ultramicroscopy}}
  \textbf{\bibinfo{volume}{108}}, \bibinfo{pages}{179--195}
  (\bibinfo{year}{2008}).
\newblock \doiprefix 10.1016/j.ultramic.2007.07.010.

\bibitem{Susi17UM}
\bibinfo{author}{Susi, T.}, \bibinfo{author}{Meyer, J.} \&
  \bibinfo{author}{Kotakoski, J.}
\newblock \bibinfo{title}{Manipulating low-dimensional materials down to the
  level of single atoms with electron irradiation}.
\newblock \emph{\bibinfo{journal}{Ultramicroscopy}}
  \textbf{\bibinfo{volume}{180}}, \bibinfo{pages}{163--172}
  (\bibinfo{year}{2017}).
\newblock \doiprefix 10.1016/j.ultramic.2017.03.005.

\bibitem{Hawkes09PTRSL}
\bibinfo{author}{Hawkes, P.~W.}
\newblock \bibinfo{title}{Aberration correction past and present}.
\newblock \emph{\bibinfo{journal}{Philosophical Transactions of the Royal
  Society of London A: Mathematical, Physical and Engineering Sciences}}
  \textbf{\bibinfo{volume}{367}}, \bibinfo{pages}{3637--3664}
  (\bibinfo{year}{2009}).
\newblock \doiprefix 10.1098/rsta.2009.0004.

\bibitem{Krivanek97Cs}
\bibinfo{author}{Krivanek, O.}, \bibinfo{author}{Dellby, N.},
  \bibinfo{author}{Spence, A.}, \bibinfo{author}{Camps, R.} \&
  \bibinfo{author}{Brown, L.}
\newblock \bibinfo{title}{{Aberration correction in the STEM}}.
\newblock In \bibinfo{editor}{Rodenburg, J.} (ed.)
  \emph{\bibinfo{booktitle}{Inst. Phys. Conf. Ser. 153 (Proceedings 1997 EMAG
  meeting)}}, \bibinfo{pages}{35--40} (\bibinfo{year}{1997}).

\bibitem{Haider98N}
\bibinfo{author}{Haider, M.} \emph{et~al.}
\newblock \bibinfo{title}{Electron microscopy image enhanced}.
\newblock \emph{\bibinfo{journal}{Nature}} \textbf{\bibinfo{volume}{392}},
  \bibinfo{pages}{768} (\bibinfo{year}{1998}).
\newblock \doiprefix 10.1038/33823.

\bibitem{Batson2002}
\bibinfo{author}{Batson, P.~E.}, \bibinfo{author}{Dellby, N.} \&
  \bibinfo{author}{Krivanek, O.~L.}
\newblock \bibinfo{title}{{Sub-{\aa}ngstrom resolution using aberration
  corrected electron optics.}}
\newblock \emph{\bibinfo{journal}{Nature}} \textbf{\bibinfo{volume}{418}},
  \bibinfo{pages}{617--20} (\bibinfo{year}{2002}).
\newblock \doiprefix 10.1038/nature00972.

\bibitem{Sawada2005}
\bibinfo{author}{Sawada, H.} \emph{et~al.}
\newblock \bibinfo{title}{{Experimental evaluation of a spherical
  aberration-corrected TEM and STEM}}.
\newblock \emph{\bibinfo{journal}{Journal of Electron Microscopy}}
  \textbf{\bibinfo{volume}{54}}, \bibinfo{pages}{119--121}
  (\bibinfo{year}{2005}).
\newblock \doiprefix 10.1093/jmicro/dfi001.

\bibitem{Dahmen2009}
\bibinfo{author}{Dahmen, U.} \emph{et~al.}
\newblock \bibinfo{title}{{Background, status and future of the Transmission
  Electron Aberration-corrected Microscope project.}}
\newblock \emph{\bibinfo{journal}{Philosophical transactions. Series A,
  Mathematical, physical, and engineering sciences}}
  \textbf{\bibinfo{volume}{367}}, \bibinfo{pages}{3795--808}
  (\bibinfo{year}{2009}).
\newblock \doiprefix 10.1098/rsta.2009.0094.

\bibitem{Kaiser11U}
\bibinfo{author}{Kaiser, U.} \emph{et~al.}
\newblock \bibinfo{title}{Transmission electron microscopy at 20 kv for imaging
  and spectroscopy}.
\newblock \emph{\bibinfo{journal}{Ultramicroscopy}}
  \textbf{\bibinfo{volume}{111}}, \bibinfo{pages}{1239--1246}
  (\bibinfo{year}{2011}).
\newblock \doiprefix 10.1016/j.ultramic.2011.03.012.

\bibitem{Hosokawa2013}
\bibinfo{author}{Hosokawa, F.}, \bibinfo{author}{Sawada, H.},
  \bibinfo{author}{Kondo, Y.}, \bibinfo{author}{Takayanagi, K.} \&
  \bibinfo{author}{Suenaga, K.}
\newblock \bibinfo{title}{{Development of Cs and Cc correctors for transmission
  electron microscopy}}.
\newblock \emph{\bibinfo{journal}{Microscopy}} \textbf{\bibinfo{volume}{62}},
  \bibinfo{pages}{23--41} (\bibinfo{year}{2013}).
\newblock \doiprefix 10.1093/jmicro/dfs134.

\bibitem{Tanaka2008}
\bibinfo{author}{Tanaka, N.}
\newblock \bibinfo{title}{{Present status and future prospects of spherical
  aberration corrected TEM/STEM for study of nanomaterials}}.
\newblock \emph{\bibinfo{journal}{Science and Technology of Advanced
  Materials}} \textbf{\bibinfo{volume}{9}}, \bibinfo{pages}{014111}
  (\bibinfo{year}{2008}).
\newblock \doiprefix 10.1088/1468-6996/9/1/014111.

\bibitem{Pennycook2011a}
\bibinfo{author}{Pennycook, S.~J.} \& \bibinfo{author}{Varela, M.}
\newblock \bibinfo{title}{{New views of materials through aberration-corrected
  scanning transmission electron microscopy}}.
\newblock \emph{\bibinfo{journal}{Microscopy}} \textbf{\bibinfo{volume}{60}},
  \bibinfo{pages}{S213--S223} (\bibinfo{year}{2011}).
\newblock \doiprefix 10.1093/jmicro/dfr030.

\bibitem{Krivanek10N}
\bibinfo{author}{Krivanek, O.~L.} \emph{et~al.}
\newblock \bibinfo{title}{Atom-by-atom structural and chemical analysis by
  annular dark-field electron microscopy}.
\newblock \emph{\bibinfo{journal}{Nature}} \textbf{\bibinfo{volume}{464}},
  \bibinfo{pages}{571--574} (\bibinfo{year}{2010}).
\newblock \doiprefix 10.1038/nature08879.

\bibitem{lovejoy_single_2012}
\bibinfo{author}{Lovejoy, T.~C.} \emph{et~al.}
\newblock \bibinfo{title}{Single atom identification by energy dispersive x-ray
  spectroscopy}.
\newblock \emph{\bibinfo{journal}{Appl. Phys. Lett.}}
  \textbf{\bibinfo{volume}{100}}, \bibinfo{pages}{154101}
  (\bibinfo{year}{2012}).
\newblock \doiprefix 10.1063/1.3701598.

\bibitem{Varela2004}
\bibinfo{author}{Varela, M.} \emph{et~al.}
\newblock \bibinfo{title}{Spectroscopic imaging of single atoms within a bulk
  solid}.
\newblock \emph{\bibinfo{journal}{Physical Review Letters}}
  \textbf{\bibinfo{volume}{92}}, \bibinfo{pages}{3--6} (\bibinfo{year}{2004}).
\newblock \doiprefix 10.1103/PhysRevLett.92.095502.

\bibitem{Muller2008}
\bibinfo{author}{Muller, D.~A.} \emph{et~al.}
\newblock \bibinfo{title}{{Atomic-scale chemical imaging of composition and
  bonding by aberration-corrected microscopy.}}
\newblock \emph{\bibinfo{journal}{Science}} \textbf{\bibinfo{volume}{319}},
  \bibinfo{pages}{1073--6} (\bibinfo{year}{2008}).
\newblock \doiprefix 10.1126/science.1148820.

\bibitem{Suenaga09NC}
\bibinfo{author}{Suenaga, K.} \emph{et~al.}
\newblock \bibinfo{title}{Visualizing and identifying single atoms using
  electron energy-loss spectroscopy with low accelerating voltage}.
\newblock \emph{\bibinfo{journal}{Nat. Chem.}} \textbf{\bibinfo{volume}{1}},
  \bibinfo{pages}{415--418} (\bibinfo{year}{2009}).
\newblock \doiprefix 10.1038/nchem.282.

\bibitem{Colliex2012a}
\bibinfo{author}{Colliex, C.} \emph{et~al.}
\newblock \bibinfo{title}{{Capturing the signature of single atoms with the
  tiny probe of a STEM.}}
\newblock \emph{\bibinfo{journal}{Ultramicroscopy}}
  \textbf{\bibinfo{volume}{123}}, \bibinfo{pages}{80--9}
  (\bibinfo{year}{2012}).
\newblock \doiprefix 10.1016/j.ultramic.2012.04.003.

\bibitem{Bangert13NL}
\bibinfo{author}{Bangert, U.} \emph{et~al.}
\newblock \bibinfo{title}{Ion implantation of graphene: Toward {IC} compatible
  technologies}.
\newblock \emph{\bibinfo{journal}{Nano Letters}} \textbf{\bibinfo{volume}{13}},
  \bibinfo{pages}{4902--4907} (\bibinfo{year}{2013}).
\newblock \doiprefix 10.1021/nl402812y.

\bibitem{Meyer12PRL}
\bibinfo{author}{Meyer, J.~C.} \emph{et~al.}
\newblock \bibinfo{title}{Accurate measurement of electron beam induced
  displacement cross sections for single-layer graphene}.
\newblock \emph{\bibinfo{journal}{Phys. Rev. Lett.}}
  \textbf{\bibinfo{volume}{108}}, \bibinfo{pages}{196102}
  (\bibinfo{year}{2012}).
\newblock \doiprefix 10.1103/PhysRevLett.108.196102.

\bibitem{Susi16NC}
\bibinfo{author}{Susi, T.} \emph{et~al.}
\newblock \bibinfo{title}{Isotope analysis in the transmission electron
  microscope}.
\newblock \emph{\bibinfo{journal}{Nature Communications}}
  \textbf{\bibinfo{volume}{7}}, \bibinfo{pages}{13040} (\bibinfo{year}{2016}).
\newblock \doiprefix 10.1038/ncomms13040.

\bibitem{Novoselov05N}
\bibinfo{author}{Novoselov, K.~S.} \emph{et~al.}
\newblock \bibinfo{title}{Two-dimensional gas of massless {D}irac fermions in
  graphene}.
\newblock \emph{\bibinfo{journal}{Nature}} \textbf{\bibinfo{volume}{438}},
  \bibinfo{pages}{197--200} (\bibinfo{year}{2005}).
\newblock \doiprefix 10.1038/nature04233.

\bibitem{Meyer08NL}
\bibinfo{author}{Meyer, J.~C.} \emph{et~al.}
\newblock \bibinfo{title}{Direct imaging of lattice atoms and topological
  defects in graphene membranes}.
\newblock \emph{\bibinfo{journal}{Nano Lett.}} \textbf{\bibinfo{volume}{8}},
  \bibinfo{pages}{3582--3586} (\bibinfo{year}{2008}).
\newblock \doiprefix 10.1021/nl801386m.

\bibitem{Geim07NM}
\bibinfo{author}{Geim, A.~K.} \& \bibinfo{author}{Novoselov, K.~S.}
\newblock \bibinfo{title}{The rise of graphene}.
\newblock \emph{\bibinfo{journal}{Nat. Mater.}} \textbf{\bibinfo{volume}{6}},
  \bibinfo{pages}{183--191} (\bibinfo{year}{2007}).
\newblock \doiprefix 10.1038/nmat1849.

\bibitem{Ramasse13NL}
\bibinfo{author}{Ramasse, Q.~M.} \emph{et~al.}
\newblock \bibinfo{title}{Probing the bonding and electronic structure of
  single atom dopants in graphene with electron energy loss spectroscopy}.
\newblock \emph{\bibinfo{journal}{Nano Letters}} \textbf{\bibinfo{volume}{13}},
  \bibinfo{pages}{4989--4995} (\bibinfo{year}{2013}).
\newblock \doiprefix 10.1021/nl304187e.

\bibitem{Zhou12PRL}
\bibinfo{author}{Zhou, W.} \emph{et~al.}
\newblock \bibinfo{title}{Direct determination of the chemical bonding of
  individual impurities in graphene}.
\newblock \emph{\bibinfo{journal}{Phys. Rev. Lett.}}
  \textbf{\bibinfo{volume}{109}}, \bibinfo{pages}{206803}
  (\bibinfo{year}{2012}).

\bibitem{Susi14PRL}
\bibinfo{author}{Susi, T.} \emph{et~al.}
\newblock \bibinfo{title}{Silicon--carbon bond inversions driven by 60-ke{V}
  electrons in graphene}.
\newblock \emph{\bibinfo{journal}{Phys. Rev. Lett.}}
  \textbf{\bibinfo{volume}{113}}, \bibinfo{pages}{115501}
  (\bibinfo{year}{2014}).
\newblock \doiprefix 10.1103/PhysRevLett.113.115501.

\bibitem{Susi172DM}
\bibinfo{author}{Susi, T.} \emph{et~al.}
\newblock \bibinfo{title}{Single-atom spectroscopy of phosphorus dopants
  implanted into graphene}.
\newblock \emph{\bibinfo{journal}{2D Materials}} \textbf{\bibinfo{volume}{4}},
  \bibinfo{pages}{021013} (\bibinfo{year}{2017}).
\newblock \doiprefix 10.1088/2053-1583/aa5e78.

\bibitem{Kotakoski14NC}
\bibinfo{author}{Kotakoski, J.}, \bibinfo{author}{Mangler, C.} \&
  \bibinfo{author}{Meyer, J.~C.}
\newblock \bibinfo{title}{Imaging atomic-level random walk of a point defect in
  graphene}.
\newblock \emph{\bibinfo{journal}{Nature Communications}}
  \textbf{\bibinfo{volume}{5}}, \bibinfo{pages}{3991} (\bibinfo{year}{2014}).
\newblock \doiprefix 10.1038/ncomms4991.

\bibitem{Kalinin16N}
\bibinfo{author}{Kalinin, S.}, \bibinfo{author}{Borisevich, A.} \&
  \bibinfo{author}{Jesse, S.}
\newblock \bibinfo{title}{Fire up the atom forge}.
\newblock \emph{\bibinfo{journal}{Nature}} \textbf{\bibinfo{volume}{539}},
  \bibinfo{pages}{485--487} (\bibinfo{year}{2016}).
\newblock \doiprefix 10.1038/539485a.

\bibitem{Plemmons15CM}
\bibinfo{author}{Plemmons, D.~A.}, \bibinfo{author}{Suri, P.~K.} \&
  \bibinfo{author}{Flannigan, D.~J.}
\newblock \bibinfo{title}{Probing structural and electronic dynamics with
  ultrafast electron microscopy}.
\newblock \emph{\bibinfo{journal}{Chemistry of Materials}}
  \textbf{\bibinfo{volume}{27}}, \bibinfo{pages}{3178--3192}
  (\bibinfo{year}{2015}).
\newblock \doiprefix 10.1021/acs.chemmater.5b00433.

\bibitem{Vogelsang15NL}
\bibinfo{author}{Vogelsang, J.} \emph{et~al.}
\newblock \bibinfo{title}{Ultrafast electron emission from a sharp metal
  nanotaper driven by adiabatic nanofocusing of surface plasmons}.
\newblock \emph{\bibinfo{journal}{Nano Letters}} \textbf{\bibinfo{volume}{15}},
  \bibinfo{pages}{4685--4691} (\bibinfo{year}{2015}).
\newblock \doiprefix 10.1021/acs.nanolett.5b01513.

\bibitem{Robertson12NC}
\bibinfo{author}{Robertson, A.~W.} \emph{et~al.}
\newblock \bibinfo{title}{Spatial control of defect creation in graphene at the
  nanoscale}.
\newblock \emph{\bibinfo{journal}{Nat. Commun.}} \textbf{\bibinfo{volume}{3}},
  \bibinfo{pages}{1144} (\bibinfo{year}{2012}).
\newblock \doiprefix 10.1038/ncomms2141.

\bibitem{Kurasch12NL}
\bibinfo{author}{Kurasch, S.} \emph{et~al.}
\newblock \bibinfo{title}{Atom-by-atom observation of grain boundary migration
  in graphene}.
\newblock \emph{\bibinfo{journal}{Nano Lett.}} \textbf{\bibinfo{volume}{12}},
  \bibinfo{pages}{3168--3173} (\bibinfo{year}{2012}).
\newblock \doiprefix 10.1021/nl301141g.

\bibitem{Girit09S}
\bibinfo{author}{Girit, {\c C}.~{\"O}.} \emph{et~al.}
\newblock \bibinfo{title}{Graphene at the edge: Stability and dynamics}.
\newblock \emph{\bibinfo{journal}{Science}} \textbf{\bibinfo{volume}{323}},
  \bibinfo{pages}{1705--1708} (\bibinfo{year}{2009}).
\newblock \doiprefix 10.1126/science.1166999.

\bibitem{Lin14NN}
\bibinfo{author}{Lin, J.} \emph{et~al.}
\newblock \bibinfo{title}{Flexible metallic nanowires with self-adaptive
  contacts to semiconducting transition-metal dichalcogenide monolayers}.
\newblock \emph{\bibinfo{journal}{Nature Nanotechnology}}
  \textbf{\bibinfo{volume}{9}}, \bibinfo{pages}{436--442}
  (\bibinfo{year}{2014}).
\newblock \doiprefix 10.1038/nnano.2014.81.

\bibitem{Monthioux14Carbon}
\bibinfo{author}{Monthioux, M.} \& \bibinfo{author}{Charlier, J.-C.}
\newblock \bibinfo{title}{Giving credit where credit is due: {The}
  {Stone}–({Thrower})–{Wales} designation revisited}.
\newblock \emph{\bibinfo{journal}{Carbon}} \textbf{\bibinfo{volume}{75}},
  \bibinfo{pages}{1--4} (\bibinfo{year}{2014}).
\newblock \doiprefix 10.1016/j.carbon.2014.03.054.

\bibitem{Wakabayashi07NN}
\bibinfo{author}{Wakabayashi, H.}, \bibinfo{author}{Koshino, M.},
  \bibinfo{author}{Sato, Y.}, \bibinfo{author}{Urita, S., K.~andIijima} \&
  \bibinfo{author}{Suenaga, K.}
\newblock \bibinfo{title}{Imaging active topological defects in carbon
  nanotubes}.
\newblock \emph{\bibinfo{journal}{Nat. Nanotechnol.}}
  \textbf{\bibinfo{volume}{2}}, \bibinfo{pages}{358} (\bibinfo{year}{2007}).
\newblock \doiprefix 10.1038/nnano.2007.141.

\bibitem{Gong14ACSNano}
\bibinfo{author}{Gong, C.} \emph{et~al.}
\newblock \bibinfo{title}{Spatially dependent lattice deformations for
  dislocations at the edges of graphene}.
\newblock \emph{\bibinfo{journal}{ACS Nano}}  (\bibinfo{year}{2014}).
\newblock \doiprefix 10.1021/nn505996c.

\bibitem{Warner12Science}
\bibinfo{author}{Warner, J.~H.} \emph{et~al.}
\newblock \bibinfo{title}{Dislocation-driven deformations in graphene}.
\newblock \emph{\bibinfo{journal}{Science}} \textbf{\bibinfo{volume}{337}},
  \bibinfo{pages}{209--212} (\bibinfo{year}{2012}).
\newblock \doiprefix 10.1126/science.1217529.

\bibitem{Lehtinen13NC}
\bibinfo{author}{Lehtinen, O.}, \bibinfo{author}{Kurasch, S.},
  \bibinfo{author}{Krasheninnikov, A.~V.} \& \bibinfo{author}{Kaiser, U.}
\newblock \bibinfo{title}{Atomic scale study of the life cycle of a dislocation
  in graphene from birth to annihilation}.
\newblock \emph{\bibinfo{journal}{Nat Commun}} \textbf{\bibinfo{volume}{4}},
  \bibinfo{pages}{2098} (\bibinfo{year}{2013}).
\newblock \doiprefix 10.1038/ncomms3098.

\bibitem{Kotakoski11PRL}
\bibinfo{author}{Kotakoski, J.}, \bibinfo{author}{Krasheninnikov, A.~V.},
  \bibinfo{author}{Kaiser, U.} \& \bibinfo{author}{Meyer, J.~C.}
\newblock \bibinfo{title}{From point defects in graphene to two-dimensional
  amorphous carbon}.
\newblock \emph{\bibinfo{journal}{Phys. Rev. Lett.}}
  \textbf{\bibinfo{volume}{106}}, \bibinfo{pages}{105505}
  (\bibinfo{year}{2011}).
\newblock \doiprefix 10.1103/PhysRevLett.106.105505.

\bibitem{Borner16PRB}
\bibinfo{author}{Börner, P.}, \bibinfo{author}{Kaiser, U.} \&
  \bibinfo{author}{Lehtinen, O.}
\newblock \bibinfo{title}{Evidence against a universal electron-beam-induced
  virtual temperature in graphene}.
\newblock \emph{\bibinfo{journal}{Phys. Rev. B}} \textbf{\bibinfo{volume}{93}},
  \bibinfo{pages}{134104} (\bibinfo{year}{2016}).
\newblock \doiprefix 10.1103/PhysRevB.93.134104.

\bibitem{Kotakoski11PRB}
\bibinfo{author}{Kotakoski, J.} \emph{et~al.}
\newblock \bibinfo{title}{Stone-{W}ales-type transformations in carbon
  nanostructures driven by electron irradiation}.
\newblock \emph{\bibinfo{journal}{Phys. Rev. B}} \textbf{\bibinfo{volume}{83}},
  \bibinfo{pages}{245420} (\bibinfo{year}{2011}).
\newblock \doiprefix 10.1103/PhysRevB.83.245420.

\bibitem{Yang14AC}
\bibinfo{author}{Yang, Z.} \emph{et~al.}
\newblock \bibinfo{title}{Direct observation of atomic dynamics and silicon
  doping at a topological defect in graphene}.
\newblock \emph{\bibinfo{journal}{Angewandte Chemie}}
  \textbf{\bibinfo{volume}{126}}, \bibinfo{pages}{9054--9058}
  (\bibinfo{year}{2014}).
\newblock \doiprefix 10.1002/ange.201403382.

\bibitem{Lee13NC}
\bibinfo{author}{Lee, J.}, \bibinfo{author}{Zhou, W.},
  \bibinfo{author}{Pennycook, S.~J.}, \bibinfo{author}{Idrobo, J.-C.} \&
  \bibinfo{author}{Pantelides, S.~T.}
\newblock \bibinfo{title}{Direct visualization of reversible dynamics in a
  {Si}$_6$ cluster embedded in a graphene pore}.
\newblock \emph{\bibinfo{journal}{Nature}} \textbf{\bibinfo{volume}{4}},
  \bibinfo{pages}{1650} (\bibinfo{year}{2013}).
\newblock \doiprefix 10.1038/ncomms2671.

\bibitem{Lin15NL}
\bibinfo{author}{Lin, Y.-C.} \emph{et~al.}
\newblock \bibinfo{title}{Structural and chemical dynamics of
  pyridinic-nitrogen defects in graphene}.
\newblock \emph{\bibinfo{journal}{Nano Letters}} \textbf{\bibinfo{volume}{15}},
  \bibinfo{pages}{7408--7413} (\bibinfo{year}{2015}).
\newblock \doiprefix 10.1021/acs.nanolett.5b02831.

\bibitem{Kepaptsoglou15AN}
\bibinfo{author}{Kepaptsoglou, D.} \emph{et~al.}
\newblock \bibinfo{title}{Electronic structure modification of ion implanted
  graphene: The spectroscopic signatures of p- and n-type doping}.
\newblock \emph{\bibinfo{journal}{ACS Nano}} \textbf{\bibinfo{volume}{9}},
  \bibinfo{pages}{11398--11407} (\bibinfo{year}{2015}).
\newblock \doiprefix 10.1021/acsnano.5b05305.

\bibitem{Skowron16C}
\bibinfo{author}{Skowron, S.~T.} \emph{et~al.}
\newblock \bibinfo{title}{Reaction kinetics of bond rotations in graphene}.
\newblock \emph{\bibinfo{journal}{Carbon}} \textbf{\bibinfo{volume}{105}},
  \bibinfo{pages}{176--182} (\bibinfo{year}{2016}).
\newblock \doiprefix 10.1016/j.carbon.2016.04.020.

\bibitem{Susi12AN}
\bibinfo{author}{Susi, T.} \emph{et~al.}
\newblock \bibinfo{title}{Atomistic description of electron beam damage in
  nitrogen-doped graphene and single-walled carbon nanotubes}.
\newblock \emph{\bibinfo{journal}{ACS Nano}} \textbf{\bibinfo{volume}{6}},
  \bibinfo{pages}{8837--8846} (\bibinfo{year}{2012}).
\newblock \doiprefix 10.1021/nn303944f.

\bibitem{Loponen06PRB}
\bibinfo{author}{Loponen, T.}, \bibinfo{author}{Krasheninnikov, A.~V.},
  \bibinfo{author}{Kaukonen, M.} \& \bibinfo{author}{Nieminen, R.~M.}
\newblock \bibinfo{title}{Nitrogen-doped carbon nanotubes under electron
  irradiation simulated with a tight-binding model}.
\newblock \emph{\bibinfo{journal}{Phys. Rev. B}} \textbf{\bibinfo{volume}{74}}
  (\bibinfo{year}{2006}).
\newblock \doiprefix 10.1103/PhysRevB.74.073409.

\bibitem{Zobelli07PRB}
\bibinfo{author}{Zobelli, A.}, \bibinfo{author}{Gloter, A.},
  \bibinfo{author}{Ewels, C.~P.}, \bibinfo{author}{Seifert, G.} \&
  \bibinfo{author}{Colliex, C.}
\newblock \bibinfo{title}{Electron knock-on cross section of carbon and boron
  nitride nanotubes}.
\newblock \emph{\bibinfo{journal}{Phys. Rev. B}} \textbf{\bibinfo{volume}{75}},
  \bibinfo{pages}{245402} (\bibinfo{year}{2007}).
\newblock \doiprefix 10.1103/PhysRevB.75.245402.

\bibitem{Kotakoski10PRB}
\bibinfo{author}{Kotakoski, J.}, \bibinfo{author}{Jin, C.},
  \bibinfo{author}{Lehtinen, O.}, \bibinfo{author}{Suenaga, K.} \&
  \bibinfo{author}{Krasheninnikov, A.}
\newblock \bibinfo{title}{Electron knock-on damage in hexagonal boron nitride
  monolayers}.
\newblock \emph{\bibinfo{journal}{Phys. Rev. B}} \textbf{\bibinfo{volume}{82}},
  \bibinfo{pages}{113404} (\bibinfo{year}{2010}).
\newblock \doiprefix 10.1103/PhysRevB.82.113404.

\bibitem{Arenal14NL}
\bibinfo{author}{Arenal, R.} \emph{et~al.}
\newblock \bibinfo{title}{Atomic configuration of nitrogen-doped single-walled
  carbon nanotubes}.
\newblock \emph{\bibinfo{journal}{Nano Letters}} \textbf{\bibinfo{volume}{14}},
  \bibinfo{pages}{5509--5516} (\bibinfo{year}{2014}).
\newblock \doiprefix 10.1021/nl501645g.

\bibitem{McKinley48PR}
\bibinfo{author}{McKinley, W.~A., Jr.} \& \bibinfo{author}{Feshbach, H.}
\newblock \bibinfo{title}{The {C}oulomb scattering of relativistic electrons by
  nuclei}.
\newblock \emph{\bibinfo{journal}{Phys. Rev.}} \textbf{\bibinfo{volume}{74}},
  \bibinfo{pages}{1759--1763} (\bibinfo{year}{1948}).
\newblock \doiprefix 10.1103/PhysRev.74.1759.

\bibitem{Perdew96PRL}
\bibinfo{author}{Perdew, J.~P.}, \bibinfo{author}{Burke, K.} \&
  \bibinfo{author}{Ernzerhof, M.}
\newblock \bibinfo{title}{Generalized gradient approximation made simple}.
\newblock \emph{\bibinfo{journal}{Phys. Rev. Lett.}}
  \textbf{\bibinfo{volume}{77}}, \bibinfo{pages}{3865--3868}
  (\bibinfo{year}{1996}).
\newblock \doiprefix 10.1103/PhysRevLett.77.3865.

\bibitem{Lippert07JCP}
\bibinfo{author}{Lippert, R.~A.} \emph{et~al.}
\newblock \bibinfo{title}{A common, avoidable source of error in molecular
  dynamics integrators}.
\newblock \emph{\bibinfo{journal}{The Journal of Chemical Physics}}
  \textbf{\bibinfo{volume}{126}}, \bibinfo{pages}{046101}
  (\bibinfo{year}{2007}).
\newblock \doiprefix 10.1063/1.2431176.

\bibitem{Egerton10UM}
\bibinfo{author}{Egerton, R.}, \bibinfo{author}{McLeod, R.},
  \bibinfo{author}{Wang, F.} \& \bibinfo{author}{Malac, M.}
\newblock \bibinfo{title}{Basic questions related to electron-induced
  sputtering in the {TEM}}.
\newblock \emph{\bibinfo{journal}{Ultramicroscopy}}
  \textbf{\bibinfo{volume}{110}}, \bibinfo{pages}{991--997}
  (\bibinfo{year}{2010}).
\newblock \doiprefix 10.1016/j.ultramic.2009.11.003.

\bibitem{Algara-Siller13APL}
\bibinfo{author}{Algara-Siller, G.}, \bibinfo{author}{Kurasch, S.},
  \bibinfo{author}{Sedighi, M.}, \bibinfo{author}{Lehtinen, O.} \&
  \bibinfo{author}{Kaiser, U.}
\newblock \bibinfo{title}{The pristine atomic structure of {M}o{S}$_2$
  monolayer protected from electron radiation damage by graphene}.
\newblock \emph{\bibinfo{journal}{Applied Physics Letters}}
  \textbf{\bibinfo{volume}{103}}, \bibinfo{pages}{203107}
  (\bibinfo{year}{2013}).
\newblock \doiprefix 10.1063/1.4830036.

\bibitem{lehnert_electron_2017}
\bibinfo{author}{Lehnert, T.}, \bibinfo{author}{Lehtinen, O.},
  \bibinfo{author}{Algara–Siller, G.} \& \bibinfo{author}{Kaiser, U.}
\newblock \bibinfo{title}{Electron radiation damage mechanisms in 2d {MoSe}2}.
\newblock \emph{\bibinfo{journal}{Appl. Phys. Lett.}}
  \textbf{\bibinfo{volume}{110}}, \bibinfo{pages}{033106}
  (\bibinfo{year}{2017}).
\newblock \doiprefix 10.1063/1.4973809.

\bibitem{Meyer08APL}
\bibinfo{author}{Meyer, J.~C.}, \bibinfo{author}{Girit, C.~O.},
  \bibinfo{author}{Crommie, M.~F.} \& \bibinfo{author}{Zettl, A.}
\newblock \bibinfo{title}{Hydrocarbon lithography on graphene membranes}.
\newblock \emph{\bibinfo{journal}{Applied Physics Letters}}
  \textbf{\bibinfo{volume}{92}}, \bibinfo{pages}{123110}
  (\bibinfo{year}{2008}).
\newblock \doiprefix 10.1063/1.2901147.

\bibitem{Lin12NL}
\bibinfo{author}{Lin, Y.-C.} \emph{et~al.}
\newblock \bibinfo{title}{Graphene annealing: How clean can it be?}
\newblock \emph{\bibinfo{journal}{Nano Letters}} \textbf{\bibinfo{volume}{12}},
  \bibinfo{pages}{414--419} (\bibinfo{year}{2012}).
\newblock \doiprefix 10.1021/nl203733r.

\bibitem{Algara-Siller14APL}
\bibinfo{author}{Algara-Siller, G.}, \bibinfo{author}{Lehtinen, O.},
  \bibinfo{author}{Turchanin, A.} \& \bibinfo{author}{Kaiser, U.}
\newblock \bibinfo{title}{Dry-cleaning of graphene}.
\newblock \emph{\bibinfo{journal}{Applied Physics Letters}}
  \textbf{\bibinfo{volume}{104}}, \bibinfo{pages}{153115}
  (\bibinfo{year}{2014}).
\newblock \doiprefix 10.1063/1.4871997.

\bibitem{Lu11NL}
\bibinfo{author}{Lu, Y.}, \bibinfo{author}{Merchant, C.~A.},
  \bibinfo{author}{Drndi{\'c}, M.} \& \bibinfo{author}{Johnson, A. T.~C.}
\newblock \bibinfo{title}{In situ electronic characterization of graphene
  nanoconstrictions fabricated in a transmission electron microscope}.
\newblock \emph{\bibinfo{journal}{Nano Letters}} \textbf{\bibinfo{volume}{11}},
  \bibinfo{pages}{5184--5188} (\bibinfo{year}{2011}).
\newblock \doiprefix 10.1021/nl2023756.

\bibitem{Tripathi17PSSRRL}
\bibinfo{author}{Tripathi, M.} \emph{et~al.}
\newblock \bibinfo{title}{Cleaning graphene: Comparing heat treatments in air
  and in vacuum}.
\newblock \emph{\bibinfo{journal}{physica status solidi (RRL) -- Rapid Research
  Letters}} \bibinfo{pages}{1700124} (\bibinfo{year}{2017}).
\newblock \doiprefix 10.1002/pssr.201700124.

\bibitem{Hong17AM}
\bibinfo{author}{Hong, J.}, \bibinfo{author}{Jin, C.}, \bibinfo{author}{Yuan,
  J.} \& \bibinfo{author}{Zhang, Z.}
\newblock \bibinfo{title}{Atomic defects in two-dimensional materials: From
  single-atom spectroscopy to functionalities in opto-/electronics,
  nanomagnetism, and catalysis}.
\newblock \emph{\bibinfo{journal}{Advanced Materials}}
  \textbf{\bibinfo{volume}{29}}, \bibinfo{pages}{1606434}
  (\bibinfo{year}{2017}).
\newblock \doiprefix 10.1002/adma.201606434.

\bibitem{Bjorkman13SciRep}
\bibinfo{author}{Björkman, T.} \emph{et~al.}
\newblock \bibinfo{title}{Defects in bilayer silica and graphene: common trends
  in diverse hexagonal two-dimensional systems}.
\newblock \emph{\bibinfo{journal}{Sci. Rep.}} \textbf{\bibinfo{volume}{3}},
  \bibinfo{pages}{3482} (\bibinfo{year}{2013}).
\newblock \doiprefix 10.1038/srep03482.

\bibitem{Gao13Nanoscale}
\bibinfo{author}{Gao, J.}, \bibinfo{author}{Zhang, J.}, \bibinfo{author}{Liu,
  H.}, \bibinfo{author}{Zhang, Q.} \& \bibinfo{author}{Zhao, J.}
\newblock \bibinfo{title}{Structures, mobilities, electronic and magnetic
  properties of point defects in silicene}.
\newblock \emph{\bibinfo{journal}{Nanoscale}} \textbf{\bibinfo{volume}{5}},
  \bibinfo{pages}{9785--9792} (\bibinfo{year}{2013}).
\newblock \doiprefix 10.1039/C3NR02826G.

\bibitem{Hu15JPCC}
\bibinfo{author}{Hu, W.} \& \bibinfo{author}{Yang, J.}
\newblock \bibinfo{title}{Defects in {Phosphorene}}.
\newblock \emph{\bibinfo{journal}{J. Phys. Chem. C}}
  \textbf{\bibinfo{volume}{119}}, \bibinfo{pages}{20474--20480}
  (\bibinfo{year}{2015}).
\newblock \doiprefix 10.1021/acs.jpcc.5b06077.

\bibitem{Komsa12PRL}
\bibinfo{author}{Komsa, H.-P.} \emph{et~al.}
\newblock \bibinfo{title}{Two-dimensional transition metal dichalcogenides
  under electron irradiation: Defect production and doping}.
\newblock \emph{\bibinfo{journal}{Phys. Rev. Lett.}}
  \textbf{\bibinfo{volume}{109}}, \bibinfo{pages}{035503}
  (\bibinfo{year}{2012}).
\newblock \doiprefix 10.1103/PhysRevLett.109.035503.

\bibitem{Guo14NC}
\bibinfo{author}{Guo, J.} \emph{et~al.}
\newblock \bibinfo{title}{Crown ethers in graphene}.
\newblock \emph{\bibinfo{journal}{Nature Communications}}
  \textbf{\bibinfo{volume}{5}}, \bibinfo{pages}{5389} (\bibinfo{year}{2014}).
\newblock \doiprefix 10.1038/ncomms6389.

\bibitem{Zhou12NN}
\bibinfo{author}{Zhou, W.} \emph{et~al.}
\newblock \bibinfo{title}{Atomically localized plasmon enhancement in monolayer
  graphene}.
\newblock \emph{\bibinfo{journal}{Nature Nanotechnology}}
  \textbf{\bibinfo{volume}{7}}, \bibinfo{pages}{161--165}
  (\bibinfo{year}{2012}).
\newblock \doiprefix 10.1038/nnano.2011.252.

\bibitem{Susi15FWF}
\bibinfo{author}{Susi, T.}
\newblock \bibinfo{title}{Heteroatom quantum corrals and nanoplasmonics in
  graphene ({HeQuCoG})}.
\newblock \emph{\bibinfo{journal}{Research Ideas and Outcomes}}
  \textbf{\bibinfo{volume}{1}}, \bibinfo{pages}{e7479} (\bibinfo{year}{2015}).
\newblock \doiprefix 10.3897/rio.1.e7479.

\end{thebibliography}

\section*{Acknowledgements}

T.S. acknowledges funding from the Austrian Science Fund (FWF) via project P~28322-N36, and the computational resources of the Vienna Scientific Cluster. D.M.K. and Q.M.R. acknowledge the support of the U.K. Engineering and Physical Sciences Research Council (EPSRC) to the SuperSTEM Laboratory, the U.K. National Facility for Aberration-Corrected STEM. Y-C.L. and K.S. acknowledge the support of JST Research Acceleration programme and JSPS KAKENHI (JP16H063333 and JP25107003). J.C.M. acknowledges funding by the European Research Council Grant No. 336453-PICOMAT. J.K. acknowledges funding from the Wiener \mbox{Wissenschafts-,} Forschungs- und Technologiefonds (WWTF) via project MA14-009 and from FWF via project I3181-N36.

\section*{Additional information}

Correspondence should be addressed to T.S. and J.K.

\section*{Competing financial interests}

The authors declare no competing financial interests.

\begin{table}[h]
\centering
\caption{\textbf{Comparison between experiment and theory for atomic displacements from graphene, its impurities, and their carbon neighbors.} Conversions between threshold energies $T_d$ (in eV) and cross sections $\sigma$ (in barn) are based on the McKinley-Feschbach approximation~\cite{McKinley48PR} with the experimental ($exp$) electron energy $E_e$. The influence of out-of-plane velocities at room temperature is included via the phonon dispersion of pristine graphene~\cite{Susi16NC}, scaled where applicable by the mass of the impurity atom. Theoretical values ($sim$) are based on DFT/MD~\cite{Susi12AN,Susi14PRL,Susi16NC} with the PBE functional~\cite{Perdew96PRL} (better match was obtained for pristine graphene with another one~\cite{Susi16NC}). No events were observed either experimentally or in the simulation when "--" is shown. The last column cites the publication describing the experiments from which the data was extracted. KO refers to a knock-on event, X-graphene refers to graphene with heteroatom dopants of element X, N$_{sub/pyr}$ refers to substitutional/pyridinic nitrogen bonding, and C@X refers to a carbon atom neighboring dopant X.}
\label{tab:thresholds}
\begin{tabular}{|l|l|l|l|l|l|l|l|l|l|l|}
\hline
System & Atom & Event & T ($^{\circ}$C) & E$_e$ (keV) & $\sigma^{exp}$ & $\sigma^{sim}$ & $T_d^{exp}$ & $T_d^{sim}$ & $\Delta T_d / T_d^{exp}$ & Ref.\\
\hline
Graphene & $^{12}$C & KO & 20 & 100 & 0.33 & 0.11 & 21.14 & 21.9 & 4\% & \cite{Susi16NC} \\
 & $^{13}$C & KO & 20 & 100 & 0.013 & 0.003 & 21.14 & 21.9 & 4\% & \cite{Susi16NC} \\
\hline
N-graphene & N$_{sub}$ & KO & 20 & 80 & $<$10$^{-9}$ & 10$^{-6}$ & $>$20.4 & 19.1 & -- & \cite{Susi12AN} \\
 & C@N$_{sub}$ & KO & 20 & 80 & 0.16 & 0.003 & 17.5 & 19.2 & 10\% & \cite{Susi12AN}\\
 & C@N$_{sub}$ & jump & 500 & 60 & 0.002 & -- & 15.9 & -- & -- & \cite{Lin15NL}\\
 & C@N$_{pyr}$ & jump & 500 & 60 & $>$30 & -- & 9.7 & -- & -- & \cite{Lin15NL}\\
\hline
B-graphene & B & KO & 20 & 60 & 0.002 & 10$^{-6}$ & 15.9 & 18 & 11\% & \cite{Kepaptsoglou15AN}\\
 & C@B & KO & 20 & 60 & 0.0009 & 10$^{-13}$ & 15.1 & 19.6 & 30\% & \cite{Kepaptsoglou15AN}\\
  & C@B & jump & 20 & 60 & 0.003 & 10$^{-10}$ & 14.8 & 18.5 & 25\% & \cite{Kepaptsoglou15AN}\\
\hline
Si-graphene & Si & KO & 20 & 60 & $<$10$^{-10}$ & 10$^{-23}$ & $>$10.0 & 13.3 & -- & \cite{Susi14PRL}\\
 & C@Si & KO & 20 & 60 & 0.02 & 10$^{-6}$ & 12.8 & 16.9 & 32\% & \cite{Susi14PRL}\\
  & C@Si & jump & 20 & 60 & 0.47 & 0.005 & 13.0 & 14.8 & 14\%& \cite{Susi14PRL}\\
\hline
\end{tabular}
\end{table}

\end{document}